# Dielectric-laser electron acceleration in a dual pillar grating with a distributed Bragg reflector

Peyman Yousefi[1,*], Norbert Schönenberger[1], Joshua McNeur[1], Martin Kozák[1,2], Uwe Niedermayer[3] and Peter Hommelhoff[1,4]

[1] *Department of Physics, Friedrich-Alexander Universität Erlangen-Nürnberg, Staudtstr. 1, 91058 Erlangen, Germany*
[2] *Faculty of Mathematics and Physics, Charles University, Ke Karlovu 3, 12116 Prague 2, Czech Republic*
[3] *Institute for Accelerator Science and Electromagnetic Fields (TEMF), Technische Universität Darmstadt, Schlossgartenstr. 8 D-64289 Darmstadt, Germany*
[4] *Max-Planck Institute for the Science of Light, Staudtstr. 2, 91058 Erlangen, Germany*

*\*Corresponding author: [peyman.yousefi@fau.de](mailto:peyman.yousefi@fau.de)*

**We report on the efficacy of a novel design for dielectric laser accelerators by adding a distributed Bragg reflector (DBR) to a dual pillar grating accelerating structure. This mimics a double-sided laser illumination, resulting in an enhanced longitudinal electric field while reducing the deflecting transverse effects, when compared to single-sided illumination. We improve the coupling efficiency of the incident electric field into the accelerating mode by 57 percent. The 12 µm long structures accelerate sub-relativistic 28 keV electrons with gradients of up to 200 MeV/m in theory and 133 MeV/m in practice. Our work shows how lithographically produced nano-structures help to make novel laser accelerators more efficient.**

Dielectric laser accelerators (DLAs) are enticing candidates for future particle accelerators. They can reduce the length of the current radio-frequency (RF) accelerators by 1 to 2 orders of magnitude as dielectrics can withstand peak electric fields in the GV/m regime and therefore yield higher acceleration gradients [1]. They are based on electrons interacting with laser-induced optical near-fields in close vicinity to a grating structure described by the inverse Smith-Purcell effect [2, 3]. Acceleration gradients of beyond 250 MeV/m using double sided fused-silica gratings at relativistic electron injection energies and up to 25 MeV/m for single sided fused-silica gratings at sub-relativistic energies have been already demonstrated [4-6]. Different structures from simple gratings to complex geometries have been studied [7-11]. Recent studies have shown acceleration gradients in excess of 200 MeV/m for 96.3 keV electrons and 210 MeV/m for 28 keV electrons using silicon structures [12, 13]. Even higher acceleration gradients of up to 370 MeV/m have been demonstrated by exciting a coupled-mode field profile using a silicon dual pillar grating for sub-100 keV electrons [14]. This design, first proposed by Palmer [15], is of interest as it can provide a practical structure for DLAs with a more uniform field profile in the acceleration channel, due to the symmetric design of the structure. This approach not only reduces the losses, dispersion and nonlinear distortion of the laser pulse, as it reduces the amount of dielectric material that the pulse needs to traverse, but also facilitates the fabrication process by eliminating the need for bonding two single sided gratings.

The uniformity in field profile is crucial for DLAs as it reduces the deflection forces for electrons [16], enabling longer accelerator structures and larger energy gains. Dual side laser illumination is a favorable approach to provide field uniformity, however it introduces more complexity in the optical path for phase matching the two laser beams. To further improve such an approach, we take advantage of a distributed Bragg reflector (DBR), a highly reflective mirror consisting of alternating layers of high and low refractive index materials, each being a quarter wavelength in thickness. Adding a DBR to the dual pillar grating mimics double-sided laser illumination, which offers a better symmetry in the transverse electric field profile between the pillars where electrons are injected. It also enhances the longitudinal electric field of the accelerating mode, which in theory can lead to an acceleration gradient as high as 70% of the incident field for 28 keV electron energies [17]. Our calculations predict a 99% reflection of the incident field from the DBR to the accelerating channel. They also show a maximum amplitude for the acceleration mode by introducing half a period ($\lambda_p/2$) offset for one row of pillars with respect to the other. This offset is found to be ideal only for low electron energies whereas for higher energies of about 100 keV, dual pillar structures without an offset yield a better efficiency of excitation for the acceleration modes [18].

To optimize the geometrical parameters of the DLA structures, we utilize a combination of FDTD field simulations [19] and General Particle Tracer (GPT) [20], a 5th order Runge-Kutta motion solver, to track the electrons through the simulated electromagnetic near-fields of the accelerator. To reduce computational requirements, the simulation is performed on a single cell with periodic boundaries in 2D, assuming the direction along the pillars' height to be invariant. This allows for quick iteration over parameter sets. Field uniformity, acceleration gradient and minimized transverse deflection are used as figures of merit for determining the ideal design. Finally, the whole structure is simulated to include edge effects. The electron beam parameters are chosen to reproduce the beam used in the experiments with energies of 28.4 keV and 28.1 keV and a geometric emittance of 300 pm rad. We assume a Gaussian beam with a FWHM circular spot size of 20 nm at the entrance of the accelerator structure.

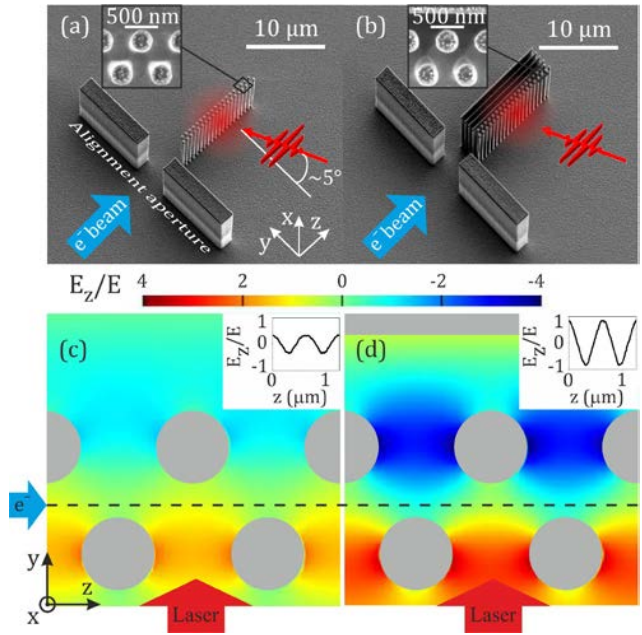

**Figure 1.** Silicon dual pillar gratings. (a) without DBR, and (b) with DBR. More details of the structures are shown in the insets. The structures are (12.0±0.1) µm long (in the z direction) and (3.0±0.1) µm tall (in the x direction). Apertures are for easier alignment of the electron beam in the channel between the pillars. The red and blue arrows indicate the directions of the laser beam and the electron beam, respectively. The laser pulse is polarized parallel to the z direction and it is incident on the structure with an angle of ~5° to the zy plane. Electron pulses are spatially and temporally superimposed with the incident laser pulses. (c) 2D-FDTD simulation of $E_z/E$ electric field profile for dual pillar grating without a DBR and (d) with a DBR, illuminated by a 1930 nm, 650 fs Gaussian laser beam. The inset plots show the cross-sectional slice of $E_z/E$ profile at the dashed line. The normalized longitudinal field amplitude is doubled when the DBR is added as the phase shift between the incident field and the reflected field from DBR is zero ($\Delta\varphi=0$).

Figure 1(a, b) shows the structures fabricated from 1-5 ohm-cm phosphorus-doped Si<100>. Their patterns are written via electron beam lithography using a RAITH 150TWO. The written patterns are etched via cryogenic reactive ion etching using a Plasma lab 100 Oxford Instruments to directionally etch silicon to a depth of (3.0±0.1) µm [21]. The structures are (12.0±0.1) µm long with a period of (630±5) nm, a pillar diameter of (320±5) nm and an acceleration channel aperture of (200±5) nm. The DBR consists of four layers of silicon with a thickness of (145±5) nm equivalent to λ/4n, with n the refractive index of silicon at the incident wavelength of λ=2 µm. The layers are separated by vacuum layers with a distance of (530±5) nm, which is approximately equal to a quarter of the designed incident wavelength. We fabricated two sets of structures one with- and one without a DBR to determine the DBR effect on the acceleration. The relative difference between geometrical dimensions of the fabricated structures is 0.8 %. Such a similarity is crucial to exclusively study the DBR effect on the electron dynamics. A 1930 nm laser pulse with a pulse duration of 650 fs is incident on the pillars with an incident angle of approximately 5° to the zy plane, exciting the optical near-fields. Alignment apertures are etched in front of the structures for easier alignment of the electron beam in the accelerating channel between the pillars.

Since the geometrical parameters of the fabricated structures deviate about 6% from the ideal simulated parameters, limited by the fabrication precision, we performed 2D-FDTD simulations for the structures with the fabricated geometrical parameters. The simulations are performed on the whole structures. Figure 1(c, d) shows the simulated longitudinal electric field ($E_z$) profile normalized by the incident electric field (E) for dual pillars without a DBR and with a DBR, respectively. A Gaussian laser beam with a central wavelength of 1930 nm and a pulse duration of 650 fs is chosen, in accordance with the values used in the experiment. It is evident from the field profiles that adding a DBR enhances the longitudinal electric field. The inset plots show the cross-sectional slice of the normalized longitudinal electric field at the dashed line. As they indicate, the field amplitude inside the accelerating channel is doubled when the DBR is added. This occurs as the phase shift between the incident field and the reflected field from the DBR is zero, $\Delta\varphi=0$, which leads to a constructive interference of the two fields.

Our experimental setup is depicted in Figure 2. We employ a Phillips XL30 scanning electron microscope (SEM) equipped with a standard Schottky emitter. A Ti:Sapphire regenerative amplifier with a repetition rate of 1 kHz, combined with an optical parametric amplifier (OPA) generates infrared (IR) pulses with a central wavelength of 1930 nm and a pulse duration of 100 fs. The IR pulses with a polarization parallel to the z axis illuminate the dual pillars from one side exciting the optical near-fields (see Figure 1d). An ultraviolet (UV) beam with a central wavelength of $\lambda_{UV}$=266 nm is generated via second harmonic generation and subsequent sum frequency generation of the fundamental and second harmonics of the Ti:Sapphire laser. The UV pulses are focused onto the Schottky tip, with a filament current below the DC emission threshold. Thus, electron pulses are emitted via single photon absorption. The resulting pulse train has an initial energy of 28 keV ($\beta$=0.32) and an initial energy spread of ~0.5 eV. After propagation through the electron optics, the electron pulses have a temporal length of $\geq$ 400 fs, measured via cross correlation with the laser pulses at the accelerator structure [22]. The electron pulses are focused to the entrance of the dual pillar channel. In order to increase the temporal overlap between the electrons and the near-fields, the IR pulses' length is stretched from 100 fs to 650 fs using a Fabry-Perot bandpass filter. This allows all electrons to interact with the excited near-fields. We temporally overlap the electron pulses with the laser illuminating the structure by controlling the arrival time of the IR pulses with a delay stage. After the electrons have interacted with the near-fields, they enter a magnetic spectrometer with a resolution of ~30 eV. The spectrally dispersed electrons are incident onto a micro-channel plate (MCP). Spectra are acquired from the phosphor screen of the MCP via a CCD camera and integrated over many iterations for each measurement to improve the signal-to-noise ratio. The spectral resolution of the detection system in the present study is limited to ∼200 eV.

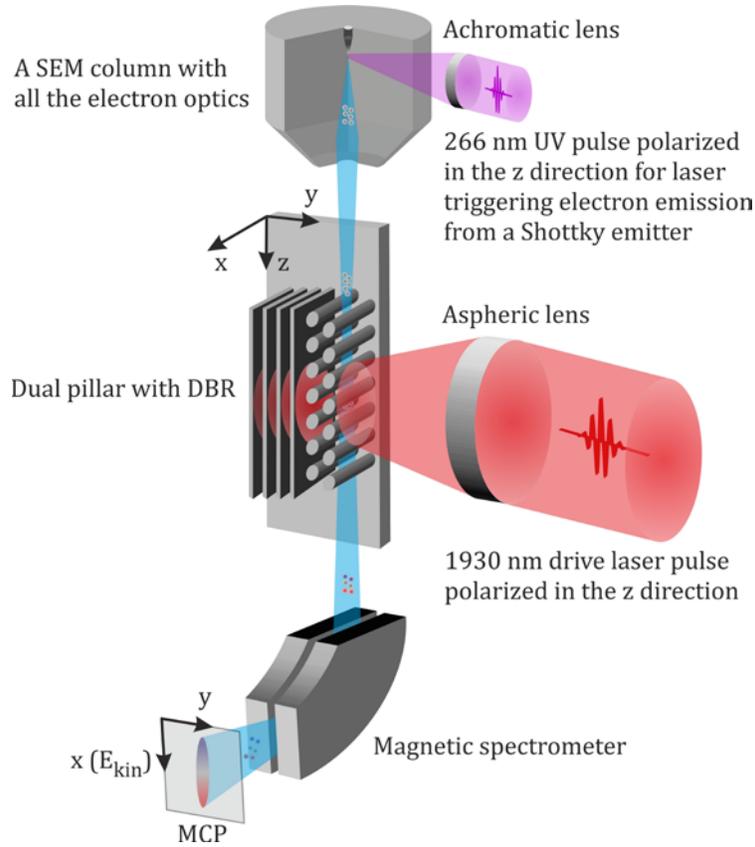

**Figure 2.** Experimental setup. Electron pulses are emitted from a UV laser-triggered Schottky emitter installed inside a SEM column. The emitted electrons traverse a dual pillar grating with DBR illuminated by 1930 nm, 650 fs drive laser pulses polarized in the z direction. After the electron pulse has interacted with the excited near-fields in between the pillars, it enters a magnetic spectrometer for energy analysis.

Figure 3 shows the measured electron energy spectra for two structures, with- and without a DBR. The peak electric field over a 9.0 µm waist radius for both structures is 0.5 GV/m. We observe a maximum energy gain of (0.44±0.05) keV for the structure without the DBR and (0.69±0.05) keV for the structure with the DBR. This corresponds to 1.57 times higher energy gain when the DBR is added. This result implies the existence of a none-zero phase shift, $\Delta\varphi \sim 0.3\pi$, between the incident field and the reflected field from the DBR. The ideal design had targeted a $\Delta\varphi=0$ to double the field amplitude in the acceleration channel. This however was not exactly achieved experimentally due to the fabrication tolerances.

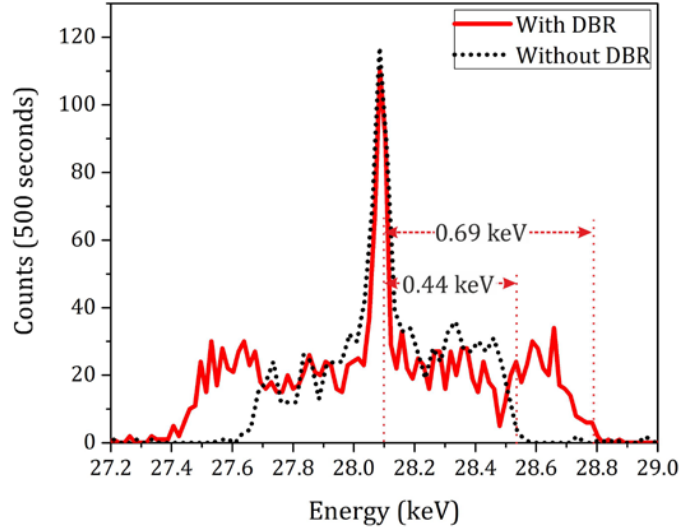

**Figure 3.** Energy spectra of 28.1 keV electrons modulated by the excited near-fields between the two rows of pillars. The dotted line shows the spectrum taken from the dual pillar grating without a DBR and the solid line denotes the structure with a DBR. Both structures are illuminated by a 1930 nm, 650 fs laser beam with a peak field amplitude of (0.50±0.1) GV/m. The structure with DBR shows 57% more energy gain for the same laser and electron beam parameters. This corresponds to a phase shift of ∼0.3π between the incident field and the reflected field from the DBR.

To measure the maximum achievable acceleration gradient, we examined a different set of dual pillar gratings with DBR whose geometrical parameters are the same as the structures used earlier. This time we used 28.4 keV electrons for different pulse energies. We observed partial structural damage at a peak electric field of (1.4±0.1) GV/m. Figure 4 shows the measured electron energy spectra for different incident peak electric fields nearly up to the structure's damage threshold. The energy spectra broaden as the peak field increases, meaning that electrons are gaining or losing more energy as the peak field increases. At 1.4 GV/m peak electric field, a maximum energy gain of (1.6±0.1) keV is achieved. The acceleration gradient ($G_{acc}$) is calculated by the longitudinal energy gain (ΔE) over the structure's length (L), $G_{acc}=\Delta E/L$. We achieved a maximum acceleration gradient of (133±9) MeV/m for a (12±0.1) µm long structure.

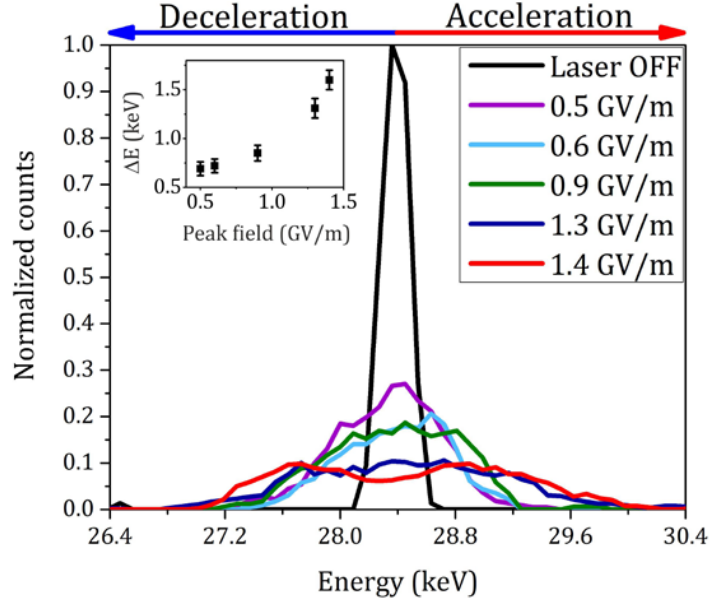

**Figure 4.** Measured energy spectra of 28.4 keV electrons after interacting with the near-fields excited by a 1930 nm laser beam with a pulse duration of 650 fs inside the acceleration channel of a dual pillar grating structure with a DBR. The maximum peak electric field of 1.4 GV/m is limited by the laser damage threshold of the structure. The inset plot shows the maximum longitudinal energy gain (ΔE) as a function of the peak electric field. A maximum energy gain of (1.6±0.1) keV is achieved for the (12±0.1) µm long structure.

To determine the efficacy of DBR in theory and the acceleration gradient limit for our current structures, we performed particle tracking simulations for structures with the same geometrical parameters as the fabricated ones. Figure 5 shows the calculated electron energy spectra as the incident peak electric field varies up to 1.4 GV/m limited by the structure's damage threshold observed experimentally. Here, 28.4 keV electrons are chosen to stay synchronized with the near-fields' phase velocity. Electrons are transmitted through the simulated electromagnetic field shown in figure 1(c, d). At 0.5 GV/m, a maximum energy gain of (0.4±0.1) keV for the structure without DBR and (0.8±0.1) keV for the structure with DBR have been calculated. The energy gain is doubled when the DBR is added, which occurs as Δφ=0 (see Figure 1(c, d)). This is an ideal condition to achieve a maximum field amplitude in the acceleration channel.

According to our simulations an energy gain of up to 2.4 keV at 1.4 GV/m peak electric field is achievable for the dual pillars with DBR, which corresponds to an acceleration gradient of 200 MeV/m for a 12 µm long structure. This is 1.5 times higher than the measured value which suggests room for further optimizations. This could be due to multiple factors such as imperfections of the fabricated structures in terms of not having perfectly round pillars or a slight alternation in the geometrical parameters from pillar to pillar. The etching process may have also roughened the pillars causing field distortions during the experiment. Moreover, the laser beam is incident with an angle which could lower the effective peak electric field in the acceleration channel, thereby lowering the energy gain.

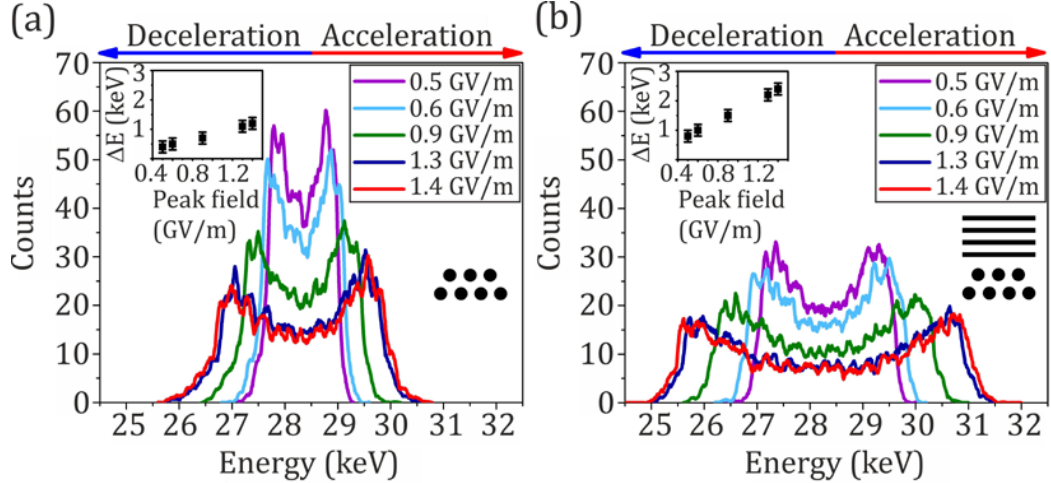

**Figure 5.** Simulated energy spectra of 28.4 keV electrons modulated by different peak electric fields for a) Dual pillars without DBR and b) Dual pillars with DBR. The peak fields are chosen in accordance to the experiment. The energy gain is doubled when the DBR is added (see the inset plots). At 1.4 GV/m a maximum energy gain of (2.4±0.1) keV is calculated which corresponds to a gradient of (200±9) MeV/m.

In conclusion, we have studied the effect of distributed Bragg reflectors on the electron energy modulation using a dual pillar grating. A DBR improves the acceleration gradient by 57% for 28 keV electrons indicating a phase shift of $\Delta\varphi \sim 0.3\pi$ between the incident field and the reflected field from the DBR. Such structures can excite a symmetric near-field profile providing up to 200 MeV/m gradient in theory and up to 133 MeV/m gradient in practice under 1930 nm, 650 fs laser illumination. Dual pillar gratings with DBR are good candidates for a miniaturized dielectric laser accelerator as they can mimic dual laser illumination on chip for a better transverse control of electrons. Since the structures are etched in one step, it offers more reproducibility in terms of geometry from structure to structure. The final DLA structure for sub-relativistic energies needs to be tapered to prevent electrons dephasing. A full particle accelerator indeed needs other elements such as pre-buncher and focusing elements [23], which can all be implemented with the DBR approach. This concept could ultimately lead to a compact laser-driven particle accelerator for variety of applications from low energy radiation therapy devices to high energy particle colliders.

**Funding.** Gordon and Betty Moore Foundation, Accelerator on a Chip International Program (ACHIP) (GBMF4744); German Federal Ministry for Education and Research (BMBF) via the project e-fs (05K16WEC); H2020 European Research Council (ERC) via the project Near Field Atto.

**Acknowledgment.** We thank F. Gannott, O. Lohse and I. Harder from the Max-Planck institute for the science of light for assistance in structure fabrication and the ACHIP collaboration members for fruitful discussions.

## References

[1] R. J. England, R. J. Noble, K. Bane, D. H. Dowell, C.-K. Ng, J. E. Spencer, S. Tantawi, Z. Wu, R. L. Byer, E. Peralta, K. Soong, C.-M. Chang, B. Montazeri, S. J. Wolf, B. Cowan, J. Dawson, W. Gai, P. Hommelhoff, Y.-C. Huang, C. Jing, C. McGuinness, R. B. Palmer, B. Naranjo, J. Rosenzweig, G. Travish, A. Mizrahi, L. Schachter, C. Sears, G. R. Werner, and R. B. Yoder, "Dielectric laser accelerators," Reviews of Modern Physics 86, 1337-1389 (2014).

[2] A. Lohmann, "Electron Acceleration by Light Waves," IBM Technical Note TN5, San Jose, CA (1962).

[3] K. Shimoda, "Proposal for an Electron Accelerator Using an Optical Maser," Appl. Opt. 1, 33-35 (1962).

[4] E. A. Peralta, K. Soong, R. J. England, E. R. Colby, Z. Wu, B. Montazeri, C. McGuinness, J. McNeur, K. J. Leedle, D. Walz, E. B. Sozer, B. Cowan, B. Schwartz, G. Travish, and R. L. Byer, "Demonstration of electron acceleration in a laser-driven dielectric microstructure," Nature 503, 91-94 (2013).

[5] J. Breuer and P. Hommelhoff, "Laser-based acceleration of nonrelativistic electrons at a dielectric structure," Physical review letters 111, 134803 (2013).

[6] D. Cesar, S. Custodio, J. Maxson, P. Musumeci, X. Shen, E. Threlkeld, R. J. England, A. Hanuka, I. V. Makasyuk, and E. A. Peralta, "High-field nonlinear optical response and phase control in a dielectric laser accelerator," Communications Physics 1, 46 (2018).

[7] C.-M. Chang and O. Solgaard, "Silicon buried gratings for dielectric laser electron accelerators," Applied Physics Letters 104, 184102 (2014).

[8] Z. Wu, R. J. England, C.-K. Ng, B. Cowan, C. McGuinness, C. Lee, M. Qi, and S. Tantawi, "Coupling power into accelerating mode of a three-dimensional silicon woodpile photonic band-gap waveguide," Physical Review Special Topics-Accelerators and Beams 17, 081301 (2014).

[9] M. Kozák, P. Beck, H. Deng, J. McNeur, N. Schönenberger, C. Gaida, F. Stutzki, M. Gebhardt, J. Limpert, and A. Ruehl, "Acceleration of sub-relativistic electrons with an evanescent optical wave at a planar interface," Optics Express 25, 19195-19204 (2017).

[10] E. A. Peralta, E. Colby, R. J. England, C. McGuinness, B. Montazeri, K. Soong, Z. Wu, and R. L. Byer, "Design, fabrication, and testing of a fused-silica dual-layer grating structure for direct laser acceleration of electrons," AIP Conference Proceedings 1507, 169-177 (2012).

[11] J. McNeur, M. Kozák, N. Schönenberger, K. J. Leedle, H. Deng, A. Ceballos, H. Hoogland, A. Ruehl, I. Hartl, and R. Holzwarth, "Elements of a dielectric laser accelerator," Optica 5, 687-690 (2018).

[12] K. J. Leedle, R. F. Pease, R. L. Byer, and J. S. Harris, "Laser acceleration and deflection of 96.3 keV electrons with a silicon dielectric structure," Optica 2, 158-161 (2015).

[13] M. Kozák, M. Förster, J. McNeur, N. Schönenberger, K. Leedle, H. Deng, J. S. Harris, R. L. Byer, and P. Hommelhoff, "Dielectric laser acceleration of sub-relativistic electrons by few-cycle laser pulses," Nuclear Instruments and Methods in Physics Research Section A: Accelerators, Spectrometers, Detectors and Associated Equipment 865, 84-86 (2017).

[14] K. J. Leedle, A. Ceballos, H. Deng, O. Solgaard, R. F. Pease, R. L. Byer, and J. S. Harris, "Dielectric laser acceleration of sub-100 keV electrons with silicon dual-pillar grating structures," Opt Lett 40, 4344-4347 (2015).

[15] R. B. Palmer, "Open accelerating structures," (CERN, 1986).

[16] J. Breuer, J. McNeur, and P. Hommelhoff, "Dielectric laser acceleration of electrons in the vicinity of single and double grating structures—theory and simulations," Journal of Physics B: Atomic, Molecular and Optical Physics 47, 234004 (2014).

[17] U. Niedermayer, O. Boine-Frankenheim, and T. Egenolf, "Designing a Dielectric Laser Accelerator on a Chip," Journal of Physics: Conference Series 874, 012041 (2017).

[18] K. J. Leedle, D. S. Black, Y. Miao, K. E. Urbanek, A. Ceballos, H. Deng, J. S. Harris, O. Solgaard, and R. L. Byer, "Phase-dependent laser acceleration of electrons with symmetrically driven silicon dual pillar gratings," Optics letters 43, 2181-2184 (2018).

[19] Lumerical FDTD Solutions, Version 8.21.1781.

[20] General Particle Tracer, Version 3.35.

[21] P. Yousefi, J. McNeur, M. Kozák, U. Niedermayer, F. Gannott, O. Lohse, O. Boine-Frankenheim, and P. Hommelhoff, "Silicon dual pillar structure with a distributed Bragg reflector for dielectric laser accelerators: Design and fabrication," Nuclear Instruments and Methods in Physics Research Section A: Accelerators, Spectrometers, Detectors and Associated Equipment 909, 221-223 (2018).

[22] M. Kozák, J. McNeur, N. Schönenberger, J. Illmer, A. Li, A. Tafel, P. Yousefi, T. Eckstein, and P. Hommelhoff, "Ultrafast scanning electron microscope applied for studying the interaction between free electrons and optical near-fields of periodic nanostructures," Journal of Applied Physics 124, 023104 (2018).

[23] U. Niedermayer, T. Egenolf, O. Boine-Frankenheim, and P. Hommelhoff, "Alternating-Phase Focusing for Dielectric-Laser Acceleration," Physical Review Letters 121, 214801 (2018).